\documentclass{article}

\usepackage{amsmath}
\usepackage{amssymb}
\usepackage{graphicx}

\usepackage{hyperref}

\usepackage{bm}
\usepackage{multirow}
\usepackage{comment}

\usepackage{amsthm}

\title{
Regularization Methods Based on the $L_q$-Likelihood
for Linear Models with Heavy-Tailed Errors\thanks{
This work was partly supported by 
JSPS KAKENHI Grant Number JP18K18008 
and JST CREST Grant Number JPMJCR1763.}
}

\author{
  Yoshihiro Hirose\thanks{hirose@ist.hokudai.ac.jp}\\
  Hokkaido University, Japan
}

\date{}

\begin{document}
\maketitle

\begin{abstract}
We propose regularization methods for linear models based on the $L_q$-likelihood, 
which is a generalization of the log-likelihood using a power function. 
Some heavy-tailed distributions are known as $q$-normal distributions. 
We find that the proposed methods for linear models with $q$-normal errors 
coincide with the regularization methods 
that are applied to the normal linear model. 
The proposed methods work well and efficiently, 
and can be computed using existing packages. 
We examine the proposed methods using numerical experiments, 
showing that the methods perform well, even when the error is 
heavy-tailed. 
\\
Keywords: 
LASSO, 
MCP, 
$q$-normal distribution, 
SCAD, 
Sparse estimation
\end{abstract}

\section{Introduction}
\label{intro}

We propose regularization methods based on the $L_q$-likelihood 
for linear models with heavy-tailed errors. 
These methods 
turn out to coincide with the ordinary regularization methods 
that are used for the normal linear model. 
The proposed methods work efficiently, 
and can be computed using existing packages. 

Linear models are widely applied, and 
many methods have been proposed for estimation, prediction, and other purposes. 
For example, for estimation and variable selection 
in the normal linear model, the literature on sparse estimation includes the 
least absolute shrinkage and selection operator (LASSO) \cite{T1996}, 
smoothly clipped absolute deviation (SCAD) \cite{FL2001}, 
Dantzig selector \cite{CT2007}, 
and minimax concave penalty (MCP) \cite{Z2010}. 
The LASSO has been studied extensively
and generalized to many models, including   
the generalized linear models \cite{PH2007}. 

Because the regularization methods for the normal linear model are useful, 
they are sometimes used in linear models with non-normal errors. 
Here, popular errors include the Cauchy error and the \textit{t}-distribution error, 
both of which are heavy-tailed errors. 
For example, \cite{AKYY2019} partly consider the Cauchy and \textit{t}-distribution errors 
in their extensive experiments. 
These heavy-tailed distributions are known to be $q$-normal distributions, 
which are studied in the literature on statistical mechanics 
\cite{F2009, PT2000, T2009}. 
The $q$-normal model is also studied 
in the literature on the generalized Cauchy distribution. 
For example, see \cite{ALFG2016, BKK2013, CAB2008, CAB2010}. 

In this study, we consider the problem of a linear regression 
with a $q$-normal error. 
We propose sparse estimation methods based on the $L_q$-likelihood, 
which is a generalization of the log-likelihood using a power function. 
The maximizer of the $L_q$-likelihood, 
the maximum $L_q$-likelihood estimator (ML$q$E), 
is investigated by \cite{FY2010} 
as an extension of the ordinary maximum likelihood estimator (MLE). 
\cite{FY2010} studies the asymptotic properties of the ML$q$E. 

The rest of the paper is organized as follows. 
In Section \ref{sec:pre}, 
we introduce several tools, including 
the normal linear model, 
regularization methods, 
$L_q$-likelihood, 
and $q$-normal models. 
In Section \ref{sec:problem}, we describe the problem under consideration,
that is, estimations in linear models with $q$-normal errors. 
Moreover, we propose several regularization methods based on the $L_q$-likelihood. 
In Section \ref{sec:numerical}, 
we evaluate the proposed methods using numerical experiments. 
Section \ref{sec:conclusion} concludes the paper.

\section{Preliminaries}
\label{sec:pre}

\subsection{Normal Linear Model and Sparse Estimation}
\label{subsec:lars}

First, we introduce the normal linear model, 
the estimation of which is a basic problem in statistics and machine learning \cite{HTF2009}. 
Furthermore, we briefly describe some well-known regularization methods. 

The normal linear model is defined as follows. 
A response is represented 
by a linear combination of explanatory variables $x_1, x_2, \dots, x_d$ as 
\begin{equation}
\label{eq:lmerror}
y^a = \theta^0 + \sum_{i=1}^d x_i^a \theta^i + \varepsilon^a \qquad(a=1,2,\dots,n), 
\end{equation}
where 
$y^a$ is the response of the $a$-th sample, 
$n$ is the sample size, 
$d$ is the number of explanatory variables, 
$x_i^a$ is the $i$-th explanatory variable of the $a$-th sample, 
$\varepsilon^a$ is a normal error with mean zero and known variance, 
and the regression coefficient ${\bm \theta}=(\theta^0, \theta^1, \dots, \theta^d)^\top$ 
is the parameter to be estimated. 
The normal linear model is equivalently given by 
\begin{equation*}
\label{eq:lm}
{\bm \mu} = X {\bm \theta}, 
\end{equation*}
where $\mu^a = \mathrm{E}[y^a]$ is the expectation of the response $y^a$, 
${\bm \mu} = (\mu^a)$, 
and $X=(x^a_i)$ is a design matrix of size $n \times (d+1)$, 
with $x^a_0=1 \, (a=1,2,\dots,n)$. 
Moreover, 
we define a row vector ${\bm x}^a \, (a=1,2,\dots,n)$ as 
${\bm x}^a = (1, x_1^a, x_2^a, \dots, x_d^a)$, 
and a column vector ${\bm x}_i \, (i=0,1,2,\dots,d)$ as 
${\bm x}_i = (x_i^1, x_i^2, \dots, x_i^n)^\top$, 
which results in 
$X = ({\bm x}^{1\top}, {\bm x}^{2\top}, \dots, {\bm x}^{n\top} )^\top 
= ({\bm x}_0, {\bm x}_1, {\bm x}_2, \dots, {\bm x}_d)$. 
Let ${\bm y} = (y^a)$ be the response vector of length $n$. 
We assume that each column vector ${\bm x}_i \, (i=1,2,\dots,d)$ is standardized, as follows: 
$\sum_{a=1}^n x_i^a = 0$ and $\| {\bm x}_i \| = 1$, for $i=1,2,\dots,d$.

As is well known, 
some regularization methods for the normal linear model 
are formulated as an optimization problem 
in the form of 
\begin{equation}
\label{eq:lasso1}
\min_{{\bm \theta} \in \mathbb{R}^{d+1}} 
\left\{ \frac{1}{2n} \| {\bm y}-X{\bm \theta} \|^2 + \rho_\lambda({\bm \theta}) \right\}, 
\end{equation}
where $\rho_\lambda({\bm \theta})$ is a penalty term, 
and $\lambda \geq 0$ is a regularization parameter. 
The LASSO \cite{T1996} uses 
$\rho_\lambda({\bm \theta}) = \lambda\| {\bm \theta} \|_1 = \lambda\sum_{i=1}^d |\theta^i|$. 
The path of the LASSO estimator when 
$\lambda$ 
varies can be made by the least angle regression (LARS) algorithm 
\cite{EHJT2004}. 
The SCAD \cite{FL2001} uses 
\begin{align}
\label{eq:scad}
\rho_\lambda({\bm \theta}) = 
\begin{cases}
\displaystyle 
\sum_{i=1}^{d} \lambda |\theta^i| \qquad (|\theta^i| \leq \lambda), \\
\displaystyle 
-\sum_{i=1}^{d} \frac{|\theta^i|^2 - 2a\lambda|\theta^i| + \lambda^2}{2(a-1)} \qquad (\lambda < |\theta^i| \leq a\lambda), \\
\displaystyle 
\sum_{i=1}^{d} \frac{(a+1)\lambda^2}{2} \qquad (a\lambda < |\theta^i|), 
\end{cases}
\end{align}
and the MCP \cite{Z2010} uses 
\begin{align}
\label{eq:mcp}
\rho_\lambda({\bm \theta}) = 
\lambda \sum_{i=1}^{d} \int_0^{|\theta^i|} \left(
1-\frac{u}{\gamma \lambda}
\right)_+ \mathrm{d}u, 
\end{align}
where $a (> 2)$ and $\gamma(>0)$ are tuning parameters. 

The regularization problem given in \eqref{eq:lasso1} can be represented by  
\begin{equation}
\label{eq:lasso2}
\min_{{\bm \theta} \in \mathbb{R}^{d+1}} 
\left\{ - \frac{1}{n} \log f({\bm y}|{\bm \theta}) + \rho_\lambda( {\bm \theta})  \right\}, 
\end{equation}
where 
$f({\bm y}|{\bm \theta})$ is the probability density function of the statistical model. 
Note that $\log f({\bm y}|{\bm \theta})$ is the log-likelihood.

\subsection{$L_q$-Likelihood}
\label{subsec:lqlike}

The $L_q$-likelihood is a generalization of the log-likelihood 
that uses a power function instead of the logarithmic function. 
Let ${\bm y}=(y^1, y^2, \dots, y^n)^\top$ be a vector of independent and identically distributed (i.i.d.) observations, 
and let $\bm \theta$ be a parameter of a statistical model.  
For $q > 0\, (q\not = 1)$, 
the $L_q$-likelihood function is defined as 
\begin{align}
\label{eq:lqlikelihood}
L_q({\bm \theta}| {\bm y}) 
= \sum_{a=1}^n \log_q f(y^a|{\bm \theta}), 
\end{align}
where $f(\cdot|{\bm\theta})$ is a probability density function of the statistical model, and 
\begin{align*}
\log_q (u) = \frac{1}{1-q}(u^{1-q} - 1) \quad (u > 0) 
\end{align*}
is the $q$-logarithmic function \cite{T2009}. 
For $q=1$, we define 
\begin{align*}
\log_1 (u) = \log u \quad (u > 0), 
\end{align*}
which is the ordinary logarithmic function. 
When $q=1$, the $L_q$-likelihood is the log-likelihood.

The ML$q$E is defined as the estimator 
that maximizes the $L_q$-likelihood. 
\cite{FY2010} studied the asymptotic performance of the ML$q$E, 
showing that it enjoys good asymptotic properties (e.g., asymptotic normality).

\subsection{$q$-Normal Model}
\label{subsec:qgauss}

Before defining the $q$-normal distribution \cite{F2009, PT2000, T2009}, 
we introduce the $q$-exponential function. 
For $q > 0\, (q\not = 1)$, 
the $q$-exponential function is the inverse function of the $q$-logarithmic function, 
and is given by 
\begin{align*}
\exp_q (u) = 
\begin{cases}
\{ 1+ (1-q)u \}^{\frac{1}{1-q}} \quad (u > -1/(1-q)), \\
0 \quad (\mathrm{otherwise}). 
\end{cases}
\end{align*}
For $q=1$, the $1$-exponential function is the ordinary exponential function  
\begin{align*}
\exp_1 (u) = \exp u \quad (u \in \mathbb{R}). 
\end{align*}

Using the $q$-exponential function, 
the $q$-normal model is given by 
\begin{align*}
{\cal S}_q &= \{ f_q(y|\, \xi, \sigma) | \xi \in \Xi, \sigma > 0 \}, \notag \\
f_q(y|\, \xi) 
&= \frac{1}{Z_q} \exp_q \left\{ -\frac{1}{3-q} \left(\frac{y - \xi}{\sigma} \right)^2 
\right\} \notag \\
&= \frac{1}{Z_q} \left\{ 1 - \frac{1-q}{3-q} \left(\frac{y - \xi}{\sigma} \right)^2 
\right\}^{\frac{1}{1-q}}, 
\end{align*}
where $\xi$ is a location parameter, 
$\Xi \subset \mathbb{R}$ is the parameter space, 
and $\sigma$ is a dispersion parameter. 
The constant $Z_q$ is a normalizing constant.   

We assume that $1 \leq q < 3$, 
which ensures that the sample space is the real line itself, 
not just part of it.  
Moreover, the parameter space is $\Xi = \mathbb{R}$ when $1 \leq q < 3$. 

For example, 
the $1$-normal model is the ordinary normal model. 
Another example is the Cauchy distribution for $q=2$: 
\begin{align*}
f_{2}(y|\mu, \sigma) 
= \frac{1}{\sigma B(\frac{1}{2}, \frac{1}{2})}\left(
1+\frac{(y-\xi)^2}{\sigma^2}
\right)^{-1}, 
\end{align*}
where $B(\cdot, \cdot)$ is the beta function. 
Furthermore, the $t$-distribution of the degree of freedom $\nu$ 
is obtained for $q=1+2/(\nu+1)$: 
\begin{align*}
f_{1+2/(\nu+1)}(y|\mu, \sigma) 
= \frac{1}{\sqrt{\nu}\sigma B(\frac{\nu}{2}, \frac{1}{2})}\left(
1+\frac{(y-\mu)^2}{\nu\sigma^2}
\right)^{-\frac{\nu+1}{2}}. 
\end{align*}

\section{Problem and Estimation Method}
\label{sec:problem}

\subsection{Linear Model with $q$-Normal Error}
\label{subsec:problem}

In this subsection, we formulate our problem, 
that is, a linear regression with a heavy-tailed error. 
The errors of the Cauchy and \textit{t}-distributions in linear models have
been studied by researchers in the context of heavy-tailed errors 
\cite{HT1975, HW1977, KM1977, S1973}. 
However, they focused mainly on the MLE, 
whereas we are interested in sparse estimators. 
Moreover, our approach is based on the $L_q$-likelihood, not the ordinary log-likelihood. 

We examine the problem of estimating the linear model given in \eqref{eq:lmerror} 
with i.i.d. errors from a $q$-normal distribution; henceforth, 
we refer to this as the $q$-normal linear model. 
In terms of probability distributions, 
we wish to estimate the parameter ${\bm \theta}$ of the $q$-normal linear model ${\cal M}_q$: 
\begin{align}
\label{eq:lqnormal}
{\cal M}_q &= \{ f(\cdot|\,{\bm \theta}) |\, {\bm \theta} \in \mathbb{R}^{d+1} \}, \notag \\
f({\bm y}|\, {\bm \theta}) 
&= \frac{1}{Z_q^n} \prod_{a=1}^n 
\exp_q \left\{ -\frac{\left(y^a - {\bm x}^a {\bm \theta} \right)^2}{3-q}  \right\} \notag \\
&= \frac{1}{Z_q^n} \prod_{a=1}^n 
\left\{ 1 - \frac{1-q}{3-q} \left(y^a - {\bm x}^a {\bm \theta} \right)^2 
\right\}^{\frac{1}{1-q}}, 
\end{align}
where the dispersion parameter is assumed to be known ($\sigma=1$). 
The $1$-normal linear model is identical to the normal linear model, as 
described in subsection \ref{subsec:lars}.

\subsection{$L_q$-likelihood-based Regularization Methods}
\label{subsec:qreguralized}

We propose regularization methods based on the $L_q$-likelihood. 
For $q$-normal linear models, 
the proposed methods coincide with  
the original regularization methods for the normal linear model. 
In other words, 
we apply the ordinary regularization methods 
as if the error distribution were a normal distribution. 
The literature describes how to compute the proposed methods efficiently. 
Moreover, 
our method calculates the ML$q$E. 

We define the $L_q$-likelihood for the $q$-normal linear model in \eqref{eq:lqnormal} 
as \eqref{eq:lqlikelihood}, 
where $\bm \theta$ is the regression coefficient. 
Note that the components of $\bm y$ are not assumed to be identically distributed 
because their distributions are dependent on the explanatory variables. 

%Thus, 
The $L_q$-likelihood for the $q$-normal linear model is 
\begin{align}
\label{eq:lq-qnormal}
L_q({\bm \theta}| {\bm y}) 
&= \sum_{a=1}^n \log_q f(y^a|{\bm \theta}) \notag \\
&= \sum_{a=1}^n \log_q \left[\frac{1}{Z_q} 
\exp_q \left\{ -\frac{\left(y^a - {\bm x}^a {\bm \theta} \right)^2}{3-q}  \right\} \right] 
\notag \\
&= -\frac{Z_q^{q-1}}{3-q} \| {\bm y}-X{\bm \theta} \|^2 - n \log_q (Z_q), 
\end{align}
where the second term is a constant. 
The ML$q$E of the parameter $\bm \theta$ is defined 
as the maximizer of the $L_q$-likelihood. 
In the $q$-normal linear model, the ML$q$E is equal to the ordinary least square, 
the MLE for the normal linear model. 

We propose a LASSO, SCAD, and MCP based on the $L_q$-likelihood 
by replacing the log-likelihood with the $L_q$-likelihood 
in the optimization problem in \eqref{eq:lasso2}. 
That is, the $L_q$-likelihood-based regularization methods are given in the form of 
\begin{equation}
\label{eq:lqlasso}
\min_{{\bm \theta} \in \mathbb{R}^{d+1}} 
\left\{ - \frac{1}{n} L_q({\bm \theta}| {\bm y}) + \rho_\lambda( {\bm \theta})  \right\}.  
\end{equation}
The penalty $\rho_\lambda$ is $\lambda\| {\bm \theta} \|_1$ for the LASSO, 
\eqref{eq:scad} for the SCAD, 
and \eqref{eq:mcp} for the MCP. 
Note that the estimator for $\lambda=0$ is the ML$q$E. 
As a special case, 
the proposed methods are the ordinary regularization methods when $q=1$. 

Because of \eqref{eq:lq-qnormal} and \eqref{eq:lqlasso}, 
for the $q$-normal linear models, 
the $L_q$-likelihood-based regularization methods are essentially the same as 
the penalized least square \eqref{eq:lasso1}. 
In other words, 
we implicitly use the $L_q$-likelihood-based regularization methods 
when we apply the ordinary LASSO, SCAD, and MCP 
to data with heavy-tailed errors.

\section{Numerical Experiments}
\label{sec:numerical}

In this section, we describe the results of our numerical experiments and compare the proposed methods. 
Here, we focus on model selection and generalization. 

Our methods do not require additional implementations 
because the LASSO, SCAD, and MCP are already implemented in software packages. 
In the experiments, 
we use the \texttt{ncvreg} package of the software R.

\subsection{Setting}
\label{subsec:setting}

The procedure for the experiments is as follows. 
We fix 
the value $q$ of the $q$-normal linear model and the $L_q$-likelihood, 
the dimension $d$ of the parameter $\bm \theta$, 
the ratio of nonzero components $r_{\mathrm{nz}}$ of $\bm \theta$, 
the true value $\theta_0$ of the nonzero components of $\bm \theta$, 
and the sample size $n$. 
The value of $q$ is selected from $1, 13/11, 3/2, 5/3, 2, 2.01, 2.1$, and $2.5$, 
where $q=13/11$ is the $t$-distribution with $\nu=10$ degrees of freedom, 
$q=3/2$ is the $t$-distribution with $\nu=3$ degrees of freedom, 
and $q=5/3$ is the $t$-distribution with $\nu=2$ degrees of freedom. 
The sample size is $n=100$ or $n=1{,}000$. 
The true parameter consists of $d \times r_{\mathrm{nz}}$ $\theta_0$s 
and $d \times (1 - r_{\mathrm{nz}})$ zeros. 
All cases are illustrated in Table \ref{tab:numerical}. 

For each of $m=1{,}000$ trials, 
we create the design matrix $X$ using the \texttt{rnorm()} function in R. 
The response $\bm y$ is generated as $q$-normal random variables 
using the \texttt{qGaussian} package. 
For the estimation, 
we apply the \texttt{ncvreg()} function to $({\bm y}, X)$ with the default options; 
for example, the values of the tuning parameters are $a=3.7$ and $\gamma=3$.

\begin{table*}%[htb]
{ 
  \begin{center}
    \caption{ 
All cases in the experiments. 
Each case is studied for the values of $q$ and $n$. 
}
   \label{tab:numerical}
%\scriptsize
%\footnotesize
\small
    \begin{tabular}{|c|c|c|c|c|c|c|c|c|} \hline
	\multirow{2}{*}{$\theta_0$} & \multicolumn{2}{|c|}{$r_{\mathrm{nz}}=0.2$} & \multicolumn{2}{|c|}{$r_{\mathrm{nz}}=0.4$} & \multicolumn{2}{|c|}{$r_{\mathrm{nz}}=0.6$} & \multicolumn{2}{|c|}{$r_{\mathrm{nz}}=0.8$} \\ \cline{2-9}
	 & $d=10$ & $d=100$ & $d=10$ & $d=100$ & $d=10$ & $d=100$ & $d=10$ & $d=100$ \\ \hline
	 $10^0$ & 1 & 5 & 9 & 13 & 17 & 21 & 25 & 29   \\ \hline %\cline{2-9}
	 $10^1$ & 2 & 6 & 10 & 14 & 18 & 22 & 26 & 30   \\ \hline %\cline{2-9}
	 $10^2$ & 3 & 7 & 11 & 15 & 19 & 23 & 27 & 31  \\ \hline
	 $10^3$ & 4 & 8 & 12 & 16 & 20 & 24 & 28 & 32  \\ \hline
    \end{tabular}
  \end{center}
}
\end{table*}

To select one model and one estimate from a sequence of parameter estimates 
generated by a method, 
we use the AIC and BIC: 
\begin{align}
\mathrm{AIC} = -2 \log p({\bm y}| \hat{{\bm \theta}}) + 2d', \label{eq:aic} \\
\mathrm{BIC} = -2 \log p({\bm y}| \hat{{\bm \theta}}) + d'\log n, \label{eq:bic}
\end{align}
where $d'$ is the dimension of parameters of the model under consideration. 
Moreover, we use other criteria based on the $L_q$-likelihood: 
\begin{align}
L_q\textrm{-AIC} = -2 L_q p({\bm y}| \hat{{\bm \theta}}) + 2d', \label{eq:lqaic} \\
L_q\textrm{-BIC} = -2 L_q p({\bm y}| \hat{{\bm \theta}}) + d'\log n. \label{eq:lqbic}
\end{align}
For a sequence $(\hat{\bm \theta}_{(k)})$ made by each of the methods, 
let $I_{(k)}=\{ i|\, \hat{\theta}_{(k)}^i \not = 0 \}$ and 
$\hat{\bm \theta}_{\mathrm{MLE}}^{(k)}$ the MLE of the model 
${\cal M}_{(k)} = \{ p(\cdot|{\bm \theta})|\, \theta^j = 0 \,\,(j \not \in I_{(k)}) \}$. 
We call 
\eqref{eq:aic} with $\hat{{\bm \theta}}=\hat{\bm \theta}_{\mathrm{MLE}}^{(k)}$ AIC1, 
and \eqref{eq:aic} with $\hat{{\bm \theta}}=\hat{\bm \theta}_{(k)}$ AIC2. 
Similarly, 
\eqref{eq:bic} with $\hat{{\bm \theta}}=\hat{\bm \theta}_{\mathrm{MLE}}^{(k)}$ is BIC1, 
and \eqref{eq:bic} with $\hat{{\bm \theta}}=\hat{\bm \theta}_{(k)}$ is BIC2. 
The $L_q$-AIC and $L_q$-BIC are referred to in the same manner; for example, 
\eqref{eq:lqaic} with $\hat{{\bm \theta}}=\hat{\bm \theta}_{\mathrm{MLE}}^{(k)}$ is $L_q$-AIC1. 
Note that AIC1, BIC1, $L_q$-AIC1, and $L_q$-BIC1 are available only when the MLE exists; 
AIC2, BIC2, $L_q$-AIC2, and $L_q$-BIC2 are always applicable. 
Finally, 
we used cross-validation (CV) 
in addition to these information criteria.

\subsection{Result}
\label{subsec:result}

The results are presented in Tables \ref{tab:model-sel_q1_n100}--\ref{tab:para-est_t2_n1000}, 
which report the best result for each method based on the various information criteria. 
%We present the complete tables of the results of the numerical experiments 
%in the Appendix. 

The model selection results are reported in Tables \ref{tab:model-sel_q1_n100}--\ref{tab:model-sel_q2p1_n1000}. 
Each number shows the number of trials (among $m=1{,}000$ trials) 
where a method selects the true model. 
Here, a larger value is better. 

The generalization results are reported in Tables \ref{tab:para-est_q1_n100}--\ref{tab:para-est_t2_n1000}. 
To evaluate the generalization error of the proposed methods, 
we newly make $m=1{,}000$ independent copies $\{ ({\bm y}'_1, X'_1), \dots, ({\bm y}'_m, X'_m) \}$ 
in each trial. 
We computed the difference between $( {\bm y}'_1, \dots, {\bm y}'_m )$ and 
the $m$ predictions using each of the methods. 
Each value shows the average prediction error over $m$ trials. 
In this case, a smaller value is better.

Our first concern is whether the proposed methods work well. 
The results for $q=1$ can be regarded as a reference for the other values of $q$. 
The tables show that the proposed methods work well in both model selection and generalization, 
especially for $q < 2$. 
The methods also perform well in terms of model selection for $q=2, 2.01$, and $2.1$. 
However, they perform poorly 
for $q=2.5$ in terms of model selection 
and for $q \geq 2$ in terms of generalization. 
As anticipated, 
a large $q$ makes the problem difficult. 

Second, we evaluate the performance of the proposed methods, finding that the MCP performs best in most cases. 
In a few cases, the MCP performed similarly to or slightly worse than the other methods. 
For model selection, 
the cases with $q=1, n=1{,}000$ and large $\theta_0$ are exceptions. 
Furthermore, the LASSO performed worse than the SCAD and MCP.  

Third, 
we consider the effect of 
$r_{\mathrm{nz}}$, 
$\theta_0$, 
$d$, 
and $n$, 
in addition to $q$. 
The cases with large $r_{\mathrm{nz}}$ 
and/or small $\theta_0$ are difficult. 
Moreover, 
a large $d$ makes the problems difficult. 
However, if we have a 
small $q$ ($1 \leq q < 2$), 
large $\theta_0$ ($\theta_0=10^2, 10^3$) 
and small $r_{\mathrm{nz}}$, 
the problems with large $d$ can be easier than those with small $d$. 
Furthermore, 
a small $n$ makes the problems difficult 
in a similar manner to a large $d$. 
These observations imply that, 
for $1 \leq q < 2$, 
small-sample problems can be easier than large-sample problems 
if $r_{\mathrm{nz}}$ is small and $\theta_0$ is large. 

Fourth, 
the choice of information criterion changes the methods' performance. 
In terms of model selection, BIC2 was mostly the best for many values of $q$. 
For $3/2 \leq q \leq 2.1$, 
BIC1 was a little better than BIC2 if BIC1 was available. 
For $q=1$ and $13/11$, 
BIC2 was better than BIC1. 
AIC1 and AIC2 were as good as BICs 
for $2 \leq q \leq 2.1$. 
Moreover, 
the $L_q$-BIC1 and -BIC2 were best only for $q=3/2$, 
when BIC1 and BIC2 performed just as well.  
Overall, the $L_q$-information criteria performed poorly. 

Furthermore, 
in terms of generalization, 
BIC2 was mostly the best. 
AIC2 was as good as BIC2, whereas AIC2 was sometimes a little worse than BIC2. 
The information criteria using the $L_q$-likelihood were poor 
for $q=13/11$. 
For $q=1, 3/2$, and $5/3$, 
the $L_q$-information criteria worked as well as the ordinary criteria and CV, 
except for some cases. 
The performance of CV was mostly good, but was occasionally very poor.

In summary, 
using an appropriate criterion, 
the proposed methods perform well for linear models with slightly heavy-tailed errors ($1 \leq q < 2$).  
Moreover, the proposed methods work in terms of model selection, even if the error is heavy-tailed ($2 \leq q < 2.5$). 
Overall, we recommend using the MCP and BIC2.

% q=1 %

\begin{table*}%[htb]
{ 
  \begin{center}
    \caption{ 
The result on model selection for $q=1$ and $n=100$. 
Each number indicates how many trials a method correctly selects the true model 
using an appropriate information criterion. 
}
   \label{tab:model-sel_q1_n100}
%\scriptsize
\footnotesize
%\small
    \begin{tabular}{|c|c|r|r|r|r|r|r|r|r|} \hline
	\multirow{2}{*}{$\theta_0$} & \multirow{2}{*}{Method} & \multicolumn{2}{|c|}{$r_{\mathrm{nz}}=0.2$} & \multicolumn{2}{|c|}{$r_{\mathrm{nz}}=0.4$} & \multicolumn{2}{|c|}{$r_{\mathrm{nz}}=0.6$} & \multicolumn{2}{|c|}{$r_{\mathrm{nz}}=0.8$} \\ \cline{3-10}
	 & & $d=10$ & $d=100$ & $d=10$ & $d=100$ & $d=10$ & $d=100$ & $d=10$ & $d=100$ \\ \hline
	 \multirow{3}{*}{$10^0$}
	  & LASSO & 32 & 0 & 9 & 0 & 5 & 0 & 8 & 0   \\ \cline{2-10} %\hline
	  & SCAD & 32 & 0 & 9 & 0 & 5 & 0 & 12 & 0   \\ \cline{2-10} %\hline
	  & MCP & 32 & 0 & 8 & 0 & 8 & 0 & 5 & 0   \\ \hline %\cline{2-10}
	 \multirow{3}{*}{$10^1$}
	  & LASSO & 366 & 0 & 486 & 0 & 625 & 0 & 784 & 0   \\ \cline{2-10} %\hline
	  & SCAD & 710 & 415 & 719 & 114 & 768 & 0 & 812 & 0   \\ \cline{2-10} %\hline
	  & MCP & 778 & 519 & 793 & 84 & 797 & 0 & 813 & 0   \\ \hline %\cline{2-10}
	 \multirow{3}{*}{$10^2$}
	  & LASSO & 366 & 35 & 487 & 0 & 627 & 0 & 788 & 0   \\ \cline{2-10} %\hline
	  & SCAD & 781 & 1{,}000 & 782 & 822 & 808 & 2 & 821 & 0   \\ \cline{2-10} %\hline
	  & MCP & 802 & 1{,}000 & 812 & 890 & 814 & 8 & 833 & 0   \\ \hline %\cline{2-10}
	 \multirow{3}{*}{$10^3$}
	  & LASSO & 365 & 46 & 496 & 0 & 655 & 0 & 823 & 0   \\ \cline{2-10} %\hline
	  & SCAD & 781 & 1{,}000 & 809 & 817 & 871 & 2 & 945 & 0   \\ \cline{2-10} %\hline
	  & MCP & 808 & 1{,}000 & 832 & 889 & 880 & 4 & 938 & 0   \\ \hline %\cline{2-10}
    \end{tabular}
  \end{center}
}
\end{table*}

\begin{table*}%[htb]
{ 
  \begin{center}
    \caption{ 
The result on model selection for $q=1$ and $n=1{,}000$. 
}
   \label{tab:model-sel_q1_n1000}
%\scriptsize
\footnotesize
%\small
    \begin{tabular}{|c|c|r|r|r|r|r|r|r|r|} \hline
	\multirow{2}{*}{$\theta_0$} & \multirow{2}{*}{Method} & \multicolumn{2}{|c|}{$r_{\mathrm{nz}}=0.2$} & \multicolumn{2}{|c|}{$r_{\mathrm{nz}}=0.4$} & \multicolumn{2}{|c|}{$r_{\mathrm{nz}}=0.6$} & \multicolumn{2}{|c|}{$r_{\mathrm{nz}}=0.8$} \\ \cline{3-10}
	 & & $d=10$ & $d=100$ & $d=10$ & $d=100$ & $d=10$ & $d=100$ & $d=10$ & $d=100$ \\ \hline
	 \multirow{3}{*}{$10^0$}
	  & LASSO & 13 & 0 & 7 & 0 & 5 & 0 & 8 & 0   \\ \cline{2-10} %\hline
	  & SCAD & 13 & 0 & 7 & 0 & 5 & 0 & 10 & 0   \\ \cline{2-10} %\hline
	  & MCP & 13 & 0 & 8 & 0 & 7 & 0 & 12 & 0   \\ \hline %\cline{2-10}
	 \multirow{3}{*}{$10^1$}
	  & LASSO & 579 & 28 & 670 & 81 & 774 & 216 & 875 & 417   \\ \cline{2-10} %\hline
	  & SCAD & 823 & 485 & 840 & 478 & 851 & 549 & 892 & 673   \\ \cline{2-10} %\hline
	  & MCP & 750 & 611 & 781 & 630 & 820 & 655 & 886 & 665   \\ \hline %\cline{2-10}
	 \multirow{3}{*}{$10^2$}
	  & LASSO & 577 & 28 & 670 & 85 & 774 & 224 & 875 & 449   \\ \cline{2-10} %\hline
	  & SCAD & 828 & 833 & 842 & 847 & 853 & 856 & 891 & 842   \\ \cline{2-10} %\hline
	  & MCP & 768 & 753 & 788 & 751 & 823 & 781 & 886 & 790   \\ \hline %\cline{2-10}
	 \multirow{3}{*}{$10^3$}
	  & LASSO & 577 & 28 & 670 & 85 & 774 & 226 & 875 & 466   \\ \cline{2-10} %\hline
	  & SCAD & 851 & 834 & 878 & 848 & 902 & 873 & 935 & 892   \\ \cline{2-10} %\hline
	  & MCP & 778 & 754 & 812 & 760 & 848 & 791 & 906 & 827   \\ \hline %\cline{2-10}
    \end{tabular}
  \end{center}
}
\end{table*}

% q=13/11 %

\begin{table*}%[htb]
{ 
  \begin{center}
    \caption{ 
The result on model selection for $q=13/11$ and $n=100$. 
}
   \label{tab:model-sel_t10_n100}
%\scriptsize
\footnotesize
%\small
    \begin{tabular}{|c|c|r|r|r|r|r|r|r|r|} \hline
	\multirow{2}{*}{$\theta_0$} & \multirow{2}{*}{Method} & \multicolumn{2}{|c|}{$r_{\mathrm{nz}}=0.2$} & \multicolumn{2}{|c|}{$r_{\mathrm{nz}}=0.4$} & \multicolumn{2}{|c|}{$r_{\mathrm{nz}}=0.6$} & \multicolumn{2}{|c|}{$r_{\mathrm{nz}}=0.8$} \\ \cline{3-10}
	 & & $d=10$ & $d=100$ & $d=10$ & $d=100$ & $d=10$ & $d=100$ & $d=10$ & $d=100$ \\ \hline
	 \multirow{3}{*}{$10^0$}
	  & LASSO & 29 & 0 & 4 & 0 & 3 & 0 & 3 & 0   \\ \cline{2-10} %\hline
	  & SCAD & 29 & 0 & 4 & 0 & 1 & 0 & 2 & 0   \\ \cline{2-10} %\hline
	  & MCP & 29 & 0 & 4 & 0 & 3 & 0 & 2 & 0   \\ \hline %\cline{2-10}
	 \multirow{3}{*}{$10^1$}
	  & LASSO & 796 & 0 & 821 & 0 & 877 & 0 & 939 & 0   \\ \cline{2-10} %\hline
	  & SCAD & 896 & 382 & 890 & 55 & 895 & 0 & 939 & 0   \\ \cline{2-10} %\hline
	  & MCP & 869 & 588 & 884 & 135 & 901 & 0 & 944 & 0   \\ \hline %\cline{2-10}
	 \multirow{3}{*}{$10^2$}
	  & LASSO & 797 & 32 & 821 & 0 & 878 & 0 & 939 & 0   \\ \cline{2-10} %\hline
	  & SCAD & 924 & 1{,}000 & 925 & 822 & 921 & 1 & 953 & 0   \\ \cline{2-10} %\hline
	  & MCP & 872 & 1{,}000 & 885 & 897 & 902 & 7 & 944 & 0   \\ \hline %\cline{2-10}
	 \multirow{3}{*}{$10^3$}
	  & LASSO & 797 & 59 & 821 & 0 & 879 & 0 & 940 & 0   \\ \cline{2-10} %\hline
	  & SCAD & 929 & 1{,}000 & 946 & 823 & 953 & 1 & 975 & 0   \\ \cline{2-10} %\hline
	  & MCP & 875 & 1{,}000 & 897 & 886 & 921 & 8 & 959 & 0   \\ \hline %\cline{2-10}
    \end{tabular}
  \end{center}
}
\end{table*}

\begin{table*}%[htb]
{ 
  \begin{center}
    \caption{ 
The result on model selection for $q=13/11$ and $n=1{,}000$. 
}
   \label{tab:model-sel_t10_n1000}
%\scriptsize
\footnotesize
%\small
    \begin{tabular}{|c|c|r|r|r|r|r|r|r|r|} \hline
	\multirow{2}{*}{$\theta_0$} & \multirow{2}{*}{Method} & \multicolumn{2}{|c|}{$r_{\mathrm{nz}}=0.2$} & \multicolumn{2}{|c|}{$r_{\mathrm{nz}}=0.4$} & \multicolumn{2}{|c|}{$r_{\mathrm{nz}}=0.6$} & \multicolumn{2}{|c|}{$r_{\mathrm{nz}}=0.8$} \\ \cline{3-10}
	 & & $d=10$ & $d=100$ & $d=10$ & $d=100$ & $d=10$ & $d=100$ & $d=10$ & $d=100$ \\ \hline
	 \multirow{3}{*}{$10^0$}
	  & LASSO & 21 & 0 & 3 & 0 & 2 & 0 & 3 & 0   \\ \cline{2-10} %\hline
	  & SCAD & 21 & 0 & 2 & 0 & 1 & 0 & 4 & 0   \\ \cline{2-10} %\hline
	  & MCP & 22 & 0 & 2 & 0 & 1 & 0 & 1 & 0   \\ \hline %\cline{2-10}
	 \multirow{3}{*}{$10^1$}
	  & LASSO & 936 & 658 & 951 & 774 & 965 & 768 & 982 & 785   \\ \cline{2-10} %\hline
	  & SCAD & 941 & 585 & 951 & 651 & 965 & 711 & 982 & 838   \\ \cline{2-10} %\hline
	  & MCP & 955 & 792 & 958 & 769 & 969 & 792 & 982 & 873   \\ \hline %\cline{2-10}
	 \multirow{3}{*}{$10^2$}
	  & LASSO & 936 & 657 & 952 & 786 & 965 & 837 & 982 & 893   \\ \cline{2-10} %\hline
	  & SCAD & 978 & 980 & 981 & 981 & 981 & 980 & 985 & 984   \\ \cline{2-10} %\hline
	  & MCP & 961 & 939 & 967 & 940 & 975 & 947 & 984 & 956   \\ \hline %\cline{2-10}
	 \multirow{3}{*}{$10^3$}
	  & LASSO & 936 & 660 & 952 & 784 & 965 & 838 & 982 & 894   \\ \cline{2-10} %\hline
	  & SCAD & 977 & 977 & 983 & 981 & 984 & 982 & 988 & 984   \\ \cline{2-10} %\hline
	  & MCP & 961 & 940 & 968 & 944 & 975 & 945 & 984 & 961   \\ \hline %\cline{2-10}
    \end{tabular}
  \end{center}
}
\end{table*}

% q=3/2 %

\begin{table*}%[htb]
{ 
  \begin{center}
    \caption{ 
The result on model selection for $q=3/2$ and $n=100$. 
}
   \label{tab:model-sel_t3_n100}
%\scriptsize
\footnotesize
%\small
    \begin{tabular}{|c|c|r|r|r|r|r|r|r|r|} \hline
	\multirow{2}{*}{$\theta_0$} & \multirow{2}{*}{Method} & \multicolumn{2}{|c|}{$r_{\mathrm{nz}}=0.2$} & \multicolumn{2}{|c|}{$r_{\mathrm{nz}}=0.4$} & \multicolumn{2}{|c|}{$r_{\mathrm{nz}}=0.6$} & \multicolumn{2}{|c|}{$r_{\mathrm{nz}}=0.8$} \\ \cline{3-10}
	 & & $d=10$ & $d=100$ & $d=10$ & $d=100$ & $d=10$ & $d=100$ & $d=10$ & $d=100$ \\ \hline
	 \multirow{3}{*}{$10^0$}
	  & LASSO & 13 & 0 & 1 & 0 & 0 & 0 & 2 & 0   \\ \cline{2-10} %\hline
	  & SCAD & 13 & 0 & 1 & 0 & 2 & 0 & 1 & 0   \\ \cline{2-10} %\hline
	  & MCP & 16 & 0 & 1 & 0 & 1 & 0 & 1 & 0   \\ \hline %\cline{2-10}
	 \multirow{3}{*}{$10^1$}
	  & LASSO & 946 & 0 & 943 & 0 & 936 & 0 & 924 & 0   \\ \cline{2-10} %\hline
	  & SCAD & 947 & 51 & 941 & 4 & 941 & 0 & 933 & 0   \\ \cline{2-10} %\hline
	  & MCP & 945 & 264 & 947 & 19 & 947 & 0 & 933 & 0   \\ \hline %\cline{2-10}
	 \multirow{3}{*}{$10^2$}
	  & LASSO & 952 & 16 & 958 & 0 & 973 & 0 & 977 & 0   \\ \cline{2-10} %\hline
	  & SCAD & 972 & 997 & 971 & 816 & 977 & 0 & 979 & 0   \\ \cline{2-10} %\hline
	  & MCP & 965 & 999 & 969 & 887 & 976 & 7 & 978 & 0   \\ \hline %\cline{2-10}
	 \multirow{3}{*}{$10^3$}
	  & LASSO & 951 & 41 & 958 & 0 & 973 & 0 & 978 & 0   \\ \cline{2-10} %\hline
	  & SCAD & 971 & 1{,}000 & 971 & 822 & 977 & 1 & 979 & 0   \\ \cline{2-10} %\hline
	  & MCP & 965 & 1{,}000 & 969 & 889 & 976 & 4 & 978 & 0   \\ \hline %\cline{2-10}
    \end{tabular}
  \end{center}
}
\end{table*}

\begin{table*}%[htb]
{ 
  \begin{center}
    \caption{ 
The result on model selection for $q=3/2$ and $n=1{,}000$. 
}
   \label{tab:model-sel_t3_n1000}
%\scriptsize
\footnotesize
%\small
    \begin{tabular}{|c|c|r|r|r|r|r|r|r|r|} \hline
	\multirow{2}{*}{$\theta_0$} & \multirow{2}{*}{Method} & \multicolumn{2}{|c|}{$r_{\mathrm{nz}}=0.2$} & \multicolumn{2}{|c|}{$r_{\mathrm{nz}}=0.4$} & \multicolumn{2}{|c|}{$r_{\mathrm{nz}}=0.6$} & \multicolumn{2}{|c|}{$r_{\mathrm{nz}}=0.8$} \\ \cline{3-10}
	 & & $d=10$ & $d=100$ & $d=10$ & $d=100$ & $d=10$ & $d=100$ & $d=10$ & $d=100$ \\ \hline
	 \multirow{3}{*}{$10^0$}
	  & LASSO & 7 & 0 & 1 & 0 & 1 & 0 & 1 & 0   \\ \cline{2-10} %\hline
	  & SCAD & 8 & 0 & 1 & 0 & 2 & 0 & 0 & 0   \\ \cline{2-10} %\hline
	  & MCP & 8 & 0 & 1 & 0 & 2 & 0 & 0 & 0   \\ \hline %\cline{2-10}
	 \multirow{3}{*}{$10^1$}
	  & LASSO & 971 & 735 & 967 & 539 & 978 & 376 & 990 & 419   \\ \cline{2-10} %\hline
	  & SCAD & 971 & 726 & 966 & 543 & 978 & 421 & 990 & 501   \\ \cline{2-10} %\hline
	  & MCP & 971 & 762 & 966 & 651 & 977 & 605 & 989 & 683   \\ \hline %\cline{2-10}
	 \multirow{3}{*}{$10^2$}
	  & LASSO & 994 & 963 & 996 & 968 & 996 & 975 & 999 & 973   \\ \cline{2-10} %\hline
	  & SCAD & 997 & 994 & 998 & 993 & 998 & 994 & 999 & 996   \\ \cline{2-10} %\hline
	  & MCP & 996 & 989 & 997 & 989 & 997 & 984 & 999 & 992   \\ \hline %\cline{2-10}
	 \multirow{3}{*}{$10^3$}
	  & LASSO & 994 & 963 & 996 & 967 & 996 & 975 & 999 & 974   \\ \cline{2-10} %\hline
	  & SCAD & 998 & 995 & 998 & 993 & 998 & 994 & 999 & 996   \\ \cline{2-10} %\hline
	  & MCP & 996 & 989 & 997 & 988 & 997 & 985 & 999 & 993   \\ \hline %\cline{2-10}
    \end{tabular}
  \end{center}
}
\end{table*}

% q=5/3 %

\begin{table*}%[htb]
{ 
  \begin{center}
    \caption{ 
The result on model selection for $q=5/3$ and $n=100$. 
}
   \label{tab:model-sel_t2_n100}
%\scriptsize
\footnotesize
%\small
    \begin{tabular}{|c|c|r|r|r|r|r|r|r|r|} \hline
	\multirow{2}{*}{$\theta_0$} & \multirow{2}{*}{Method} & \multicolumn{2}{|c|}{$r_{\mathrm{nz}}=0.2$} & \multicolumn{2}{|c|}{$r_{\mathrm{nz}}=0.4$} & \multicolumn{2}{|c|}{$r_{\mathrm{nz}}=0.6$} & \multicolumn{2}{|c|}{$r_{\mathrm{nz}}=0.8$} \\ \cline{3-10}
	 & & $d=10$ & $d=100$ & $d=10$ & $d=100$ & $d=10$ & $d=100$ & $d=10$ & $d=100$ \\ \hline
	 \multirow{3}{*}{$10^0$}
	  & LASSO & 4 & 0 & 1 & 0 & 3 & 0 & 11 & 0   \\ \cline{2-10} %\hline
	  & SCAD & 5 & 0 & 1 & 0 & 3 & 0 & 9 & 0   \\ \cline{2-10} %\hline
	  & MCP & 4 & 0 & 1 & 0 & 3 & 0 & 5 & 0   \\ \hline %\cline{2-10}
	 \multirow{3}{*}{$10^1$}
	  & LASSO & 779 & 0 & 727 & 0 & 722 & 0 & 762 & 0   \\ \cline{2-10} %\hline
	  & SCAD & 785 & 1 & 728 & 0 & 722 & 0 & 764 & 0   \\ \cline{2-10} %\hline
	  & MCP & 791 & 12 & 739 & 0 & 745 & 0 & 779 & 0   \\ \hline %\cline{2-10}
	 \multirow{3}{*}{$10^2$}
	  & LASSO & 929 & 11 & 935 & 0 & 932 & 0 & 947 & 0   \\ \cline{2-10} %\hline
	  & SCAD & 953 & 955 & 950 & 770 & 945 & 1 & 952 & 0   \\ \cline{2-10} %\hline
	  & MCP & 933 & 956 & 931 & 857 & 935 & 9 & 949 & 0   \\ \hline %\cline{2-10}
	 \multirow{3}{*}{$10^3$}
	  & LASSO & 930 & 50 & 933 & 0 & 935 & 0 & 948 & 0   \\ \cline{2-10} %\hline
	  & SCAD & 954 & 1{,}000 & 951 & 815 & 949 & 0 & 964 & 0   \\ \cline{2-10} %\hline
	  & MCP & 954 & 1{,}000 & 934 & 893 & 937 & 3 & 959 & 0   \\ \hline %\cline{2-10}
    \end{tabular}
  \end{center}
}
\end{table*}

\begin{table*}%[htb]
{ 
  \begin{center}
    \caption{ 
The result on model selection for $q=5/3$ and $n=1{,}000$. 
}
   \label{tab:model-sel_t2_n1000}
%\scriptsize
\footnotesize
%\small
    \begin{tabular}{|c|c|r|r|r|r|r|r|r|r|} \hline
	\multirow{2}{*}{$\theta_0$} & \multirow{2}{*}{Method} & \multicolumn{2}{|c|}{$r_{\mathrm{nz}}=0.2$} & \multicolumn{2}{|c|}{$r_{\mathrm{nz}}=0.4$} & \multicolumn{2}{|c|}{$r_{\mathrm{nz}}=0.6$} & \multicolumn{2}{|c|}{$r_{\mathrm{nz}}=0.8$} \\ \cline{3-10}
	 & & $d=10$ & $d=100$ & $d=10$ & $d=100$ & $d=10$ & $d=100$ & $d=10$ & $d=100$ \\ \hline
	 \multirow{3}{*}{$10^0$}
	  & LASSO & 2 & 0 & 2 & 0 & 2 & 0 & 4 & 0   \\ \cline{2-10} %\hline
	  & SCAD & 2 & 0 & 2 & 0 & 2 & 0 & 5 & 0   \\ \cline{2-10} %\hline
	  & MCP & 2 & 0 & 1 & 0 & 2 & 0 & 4 & 0   \\ \hline %\cline{2-10}
	 \multirow{3}{*}{$10^1$}
	  & LASSO & 810 & 87 & 750 & 22 & 729 & 18 & 794 & 35   \\ \cline{2-10} %\hline
	  & SCAD & 810 & 90 & 750 & 27 & 725 & 27 & 798 & 28   \\ \cline{2-10} %\hline
	  & MCP & 810 & 106 & 749 & 48 & 731 & 41 & 799 & 58   \\ \hline %\cline{2-10}
	 \multirow{3}{*}{$10^2$}
	  & LASSO & 981 & 984 & 984 & 984 & 986 & 975 & 994 & 956   \\ \cline{2-10} %\hline
	  & SCAD & 981 & 987 & 984 & 983 & 986 & 977 & 994 & 971   \\ \cline{2-10} %\hline
	  & MCP & 983 & 987 & 984 & 982 & 986 & 979 & 994 & 972   \\ \hline %\cline{2-10}
	 \multirow{3}{*}{$10^3$}
	  & LASSO & 983 & 988 & 986 & 987 & 986 & 981 & 995 & 969   \\ \cline{2-10} %\hline
	  & SCAD & 986 & 989 & 989 & 983 & 990 & 981 & 995 & 974   \\ \cline{2-10} %\hline
	  & MCP & 983 & 987 & 986 & 983 & 987 & 981 & 995 & 974   \\ \hline %\cline{2-10}
    \end{tabular}
  \end{center}
}
\end{table*}

% q=2 %

\begin{table*}%[htb]
{ 
  \begin{center}
    \caption{ 
The result on model selection for $q=2$ and $n=100$. 
}
   \label{tab:model-sel_q2_n100}
%\scriptsize
\footnotesize
%\small
    \begin{tabular}{|c|c|r|r|r|r|r|r|r|r|} \hline
	\multirow{2}{*}{$\theta_0$} & \multirow{2}{*}{Method} & \multicolumn{2}{|c|}{$r_{\mathrm{nz}}=0.2$} & \multicolumn{2}{|c|}{$r_{\mathrm{nz}}=0.4$} & \multicolumn{2}{|c|}{$r_{\mathrm{nz}}=0.6$} & \multicolumn{2}{|c|}{$r_{\mathrm{nz}}=0.8$} \\ \cline{3-10}
	 & & $d=10$ & $d=100$ & $d=10$ & $d=100$ & $d=10$ & $d=100$ & $d=10$ & $d=100$ \\ \hline
	 \multirow{3}{*}{$10^0$}
	  & LASSO & 3 & 0 & 2 & 0 & 0 & 0 & 2 & 0   \\ \cline{2-10} %\hline
	  & SCAD & 3 & 0 & 2 & 0 & 0 & 0 & 2 & 0   \\ \cline{2-10} %\hline
	  & MCP & 3 & 0 & 1 & 0 & 0 & 0 & 2 & 0   \\ \hline %\cline{2-10}
	 \multirow{3}{*}{$10^1$}
	  & LASSO & 81 & 0 & 57 & 0 & 53 & 0 & 58 & 0   \\ \cline{2-10} %\hline
	  & SCAD & 83 & 0 & 58 & 0 & 50 & 0 & 54 & 0   \\ \cline{2-10} %\hline
	  & MCP & 83 & 0 & 57 & 0 & 50 & 0 & 57 & 0   \\ \hline %\cline{2-10}
	 \multirow{3}{*}{$10^2$}
	  & LASSO & 821 & 0 &777 & 0 & 745 & 0 & 732 & 0   \\ \cline{2-10} %\hline
	  & SCAD & 823 & 314 & 782 & 219 & 755 & 2 & 735 & 0   \\ \cline{2-10} %\hline
	  & MCP & 824 & 357 & 786 & 229 & 756 & 0 & 736 & 0   \\ \hline %\cline{2-10}
	 \multirow{3}{*}{$10^3$}
	  & LASSO & 961 & 29 & 935 & 0 & 895 & 0 & 877 & 0   \\ \cline{2-10} %\hline
	  & SCAD & 966 & 915 & 944 & 746 & 913 & 1 & 879 & 0   \\ \cline{2-10} %\hline
	  & MCP & 967 & 935 & 945 & 820 & 913 & 7 & 879 & 0   \\ \hline %\cline{2-10}
    \end{tabular}
  \end{center}
}
\end{table*}

\begin{table*}%[htb]
{ 
  \begin{center}
    \caption{ 
The result on model selection for $q=2$ and $n=1{,}000$. 
}
   \label{tab:model-sel_q2_n1000}
%\scriptsize
\footnotesize
%\small
    \begin{tabular}{|c|c|r|r|r|r|r|r|r|r|} \hline
	\multirow{2}{*}{$\theta_0$} & \multirow{2}{*}{Method} & \multicolumn{2}{|c|}{$r_{\mathrm{nz}}=0.2$} & \multicolumn{2}{|c|}{$r_{\mathrm{nz}}=0.4$} & \multicolumn{2}{|c|}{$r_{\mathrm{nz}}=0.6$} & \multicolumn{2}{|c|}{$r_{\mathrm{nz}}=0.8$} \\ \cline{3-10}
	 & & $d=10$ & $d=100$ & $d=10$ & $d=100$ & $d=10$ & $d=100$ & $d=10$ & $d=100$ \\ \hline
	 \multirow{3}{*}{$10^0$}
	  & LASSO & 1 & 0 & 1 & 0 & 0 & 0 & 3 & 0   \\ \cline{2-10} %\hline
	  & SCAD & 3 & 0 & 1 & 0 & 0 & 0 & 3 & 0   \\ \cline{2-10} %\hline
	  & MCP & 2 & 0 & 0 & 0 & 0 & 0 & 3 & 0   \\ \hline %\cline{2-10}
	 \multirow{3}{*}{$10^1$}
	  & LASSO & 5 & 0 & 1 & 0 & 1 & 0 & 1 & 0   \\ \cline{2-10} %\hline
	  & SCAD & 5 & 0 & 1 & 0 & 1 & 0 & 1 & 0   \\ \cline{2-10} %\hline
	  & MCP & 4 & 0 & 0 & 0 & 0 & 0 & 1 & 0   \\ \hline %\cline{2-10}
	 \multirow{3}{*}{$10^2$}
	  & LASSO & 518 & 221 & 476 & 162 & 451 & 132 & 494 & 123   \\ \cline{2-10} %\hline
	  & SCAD & 519 & 226 & 475 & 162 & 453 & 142 & 494 & 145   \\ \cline{2-10} %\hline
	  & MCP & 518 & 235 & 473 & 200 & 455 & 180 & 495 & 178   \\ \hline %\cline{2-10}
	 \multirow{3}{*}{$10^3$}
	  & LASSO & 934 & 910 & 936 & 883 & 941 & 843 & 926 & 779   \\ \cline{2-10} %\hline
	  & SCAD & 933 & 910 & 938 & 892 & 942 & 871 & 925 & 840   \\ \cline{2-10} %\hline
	  & MCP & 933 & 911 & 938 & 899 & 943 & 884 & 926 & 855   \\ \hline %\cline{2-10}
    \end{tabular}
  \end{center}
}
\end{table*}

% q=2.01 %

\begin{table*}%[htb]
{ 
  \begin{center}
    \caption{ 
The result on model selection for $q=2.01$ and $n=100$. 
}
   \label{tab:model-sel_q2p01_n100}
%\scriptsize
\footnotesize
%\small
    \begin{tabular}{|c|c|r|r|r|r|r|r|r|r|} \hline
	\multirow{2}{*}{$\theta_0$} & \multirow{2}{*}{Method} & \multicolumn{2}{|c|}{$r_{\mathrm{nz}}=0.2$} & \multicolumn{2}{|c|}{$r_{\mathrm{nz}}=0.4$} & \multicolumn{2}{|c|}{$r_{\mathrm{nz}}=0.6$} & \multicolumn{2}{|c|}{$r_{\mathrm{nz}}=0.8$} \\ \cline{3-10}
	 & & $d=10$ & $d=100$ & $d=10$ & $d=100$ & $d=10$ & $d=100$ & $d=10$ & $d=100$ \\ \hline
	 \multirow{3}{*}{$10^0$}
	  & LASSO & 1 & 0 & 0 & 0 & 0 & 0 & 1 & 0   \\ \cline{2-10} %\hline
	  & SCAD & 2 & 0 & 0 & 0 & 0 & 0 & 1 & 0   \\ \cline{2-10} %\hline
	  & MCP & 1 & 0 & 0 & 0 & 0 & 0 & 1 & 0   \\ \hline %\cline{2-10}
	 \multirow{3}{*}{$10^1$}
	  & LASSO & 84 & 0 & 44 & 0 & 37 & 0 & 48 & 0   \\ \cline{2-10} %\hline
	  & SCAD & 86 & 0 & 45 & 0 & 40 & 0 & 49 & 0   \\ \cline{2-10} %\hline
	  & MCP & 87 & 0 & 46 & 0 & 43 & 0 & 52 & 0   \\ \hline %\cline{2-10}
	 \multirow{3}{*}{$10^2$}
	  & LASSO & 834 & 1 & 793 & 0 & 775 & 0 & 773 & 0   \\ \cline{2-10} %\hline
	  & SCAD & 835 & 338 & 792 & 210 & 774 & 0 & 772 & 0   \\ \cline{2-10} %\hline
	  & MCP & 834 & 377 & 799 & 232 & 788 & 3 & 776 & 0   \\ \hline %\cline{2-10}
	 \multirow{3}{*}{$10^3$}
	  & LASSO & 971 & 22 & 961 & 0 & 937 & 0 & 921 & 0   \\ \cline{2-10} %\hline
	  & SCAD & 972 & 894 & 961 & 737 & 936 & 1 & 927 & 0   \\ \cline{2-10} %\hline
	  & MCP & 973 & 916 & 960 & 810 & 939 & 3 & 930 & 0   \\ \hline %\cline{2-10}
    \end{tabular}
  \end{center}
}
\end{table*}

\begin{table*}%[htb]
{ 
  \begin{center}
    \caption{ 
The result on model selection for $q=2.01$ and $n=1{,}000$. 
}
   \label{tab:model-sel_q2p01_n1000}
%\scriptsize
\footnotesize
%\small
    \begin{tabular}{|c|c|r|r|r|r|r|r|r|r|} \hline
	\multirow{2}{*}{$\theta_0$} & \multirow{2}{*}{Method} & \multicolumn{2}{|c|}{$r_{\mathrm{nz}}=0.2$} & \multicolumn{2}{|c|}{$r_{\mathrm{nz}}=0.4$} & \multicolumn{2}{|c|}{$r_{\mathrm{nz}}=0.6$} & \multicolumn{2}{|c|}{$r_{\mathrm{nz}}=0.8$} \\ \cline{3-10}
	 & & $d=10$ & $d=100$ & $d=10$ & $d=100$ & $d=10$ & $d=100$ & $d=10$ & $d=100$ \\ \hline
	 \multirow{3}{*}{$10^0$}
	  & LASSO & 2 & 0 & 0 & 0 & 0 & 0 & 1 & 0   \\ \cline{2-10} %\hline
	  & SCAD & 3 & 0 & 0 & 0 & 0 & 0 & 1 & 0   \\ \cline{2-10} %\hline
	  & MCP & 1 & 0 & 0 & 0 & 0 & 0 & 1 & 0   \\ \hline %\cline{2-10}
	 \multirow{3}{*}{$10^1$}
	  & LASSO & 6 & 0 & 1 & 0 & 1 & 0 & 3 & 0   \\ \cline{2-10} %\hline
	  & SCAD & 6 & 0 & 1 & 0 & 1 & 0 & 3 & 0   \\ \cline{2-10} %\hline
	  & MCP & 2 & 0 & 0 & 0 & 1 & 0 & 3 & 0   \\ \hline %\cline{2-10}
	 \multirow{3}{*}{$10^2$}
	  & LASSO & 493 & 161 & 447 & 124 & 441 & 92 & 466 & 85   \\ \cline{2-10} %\hline
	  & SCAD & 493 & 165 & 445 & 123 & 442 & 94 & 464 & 107   \\ \cline{2-10} %\hline
	  & MCP & 495 & 174 & 443 & 144 & 445 & 134 & 463 & 143   \\ \hline %\cline{2-10}
	 \multirow{3}{*}{$10^3$}
	  & LASSO & 931 & 871 & 928 & 839 & 933 & 808 & 922 & 757   \\ \cline{2-10} %\hline
	  & SCAD & 931 & 872 & 929 & 846 & 935 & 826 & 920 & 815   \\ \cline{2-10} %\hline
	  & MCP & 931 & 875 & 929 & 861 & 935 & 844 & 921 & 827   \\ \hline %\cline{2-10}
    \end{tabular}
  \end{center}
}
\end{table*}

% q=2.1 %

\begin{table*}%[htb]
{ 
  \begin{center}
    \caption{ 
The result on model selection for $q=2.1$ and $n=100$. 
}
   \label{tab:model-sel_q2p1_n100}
%\scriptsize
\footnotesize
%\small
    \begin{tabular}{|c|c|r|r|r|r|r|r|r|r|} \hline
	\multirow{2}{*}{$\theta_0$} & \multirow{2}{*}{Method} & \multicolumn{2}{|c|}{$r_{\mathrm{nz}}=0.2$} & \multicolumn{2}{|c|}{$r_{\mathrm{nz}}=0.4$} & \multicolumn{2}{|c|}{$r_{\mathrm{nz}}=0.6$} & \multicolumn{2}{|c|}{$r_{\mathrm{nz}}=0.8$} \\ \cline{3-10}
	 & & $d=10$ & $d=100$ & $d=10$ & $d=100$ & $d=10$ & $d=100$ & $d=10$ & $d=100$ \\ \hline
	 \multirow{3}{*}{$10^0$}
	  & LASSO & 2 & 0 & 0 & 0 & 0 & 0 & 3 & 0   \\ \cline{2-10} %\hline
	  & SCAD & 2 & 0 & 0 & 0 & 0 & 0 & 3 & 0   \\ \cline{2-10} %\hline
	  & MCP & 3 & 0 & 0 & 0 & 0 & 0 & 3 & 0   \\ \hline %\cline{2-10}
	 \multirow{3}{*}{$10^1$}
	  & LASSO & 6 & 0 & 9 & 0 & 4 & 0 & 6 & 0   \\ \cline{2-10} %\hline
	  & SCAD & 6 & 0 & 9 & 0 & 4 & 0 & 4 & 0   \\ \cline{2-10} %\hline
	  & MCP & 4 & 0 & 10 & 0 & 5 & 0 & 5 & 0   \\ \hline %\cline{2-10}
	 \multirow{3}{*}{$10^2$}
	  & LASSO & 619 & 0 & 553 & 0 & 512 & 0 & 545 & 0   \\ \cline{2-10} %\hline
	  & SCAD & 618 & 117 & 554 & 53 & 525 & 0 & 535 & 0   \\ \cline{2-10} %\hline
	  & MCP & 621 & 139 & 564 & 58 & 537 & 0 & 545 & 0   \\ \hline %\cline{2-10}
	 \multirow{3}{*}{$10^3$}
	  & LASSO & 903 & 9 & 881 & 0 & 864 & 0 & 854 & 0   \\ \cline{2-10} %\hline
	  & SCAD & 903 & 753 & 890 & 612 & 866 & 1 & 846 & 0   \\ \cline{2-10} %\hline
	  & MCP & 904 & 778 & 892 & 652 & 871 & 3 & 847 & 0   \\ \hline %\cline{2-10}
    \end{tabular}
  \end{center}
}
\end{table*}

\begin{table*}%[htb]
{ 
  \begin{center}
    \caption{ 
The result on model selection for $q=2.1$ and $n=1{,}000$. 
}
   \label{tab:model-sel_q2p1_n1000}
%\scriptsize
\footnotesize
%\small
    \begin{tabular}{|c|c|r|r|r|r|r|r|r|r|} \hline
	\multirow{2}{*}{$\theta_0$} & \multirow{2}{*}{Method} & \multicolumn{2}{|c|}{$r_{\mathrm{nz}}=0.2$} & \multicolumn{2}{|c|}{$r_{\mathrm{nz}}=0.4$} & \multicolumn{2}{|c|}{$r_{\mathrm{nz}}=0.6$} & \multicolumn{2}{|c|}{$r_{\mathrm{nz}}=0.8$} \\ \cline{3-10}
	 & & $d=10$ & $d=100$ & $d=10$ & $d=100$ & $d=10$ & $d=100$ & $d=10$ & $d=100$ \\ \hline
	 \multirow{3}{*}{$10^0$}
	  & LASSO & 6 & 0 & 1 & 0 & 0 & 0 & 0 & 0   \\ \cline{2-10} %\hline
	  & SCAD & 6 & 0 & 1 & 0 & 0 & 0 & 0 & 0   \\ \cline{2-10} %\hline
	  & MCP & 4 & 0 & 1 & 0 & 0 & 0 & 0 & 0   \\ \hline %\cline{2-10}
	 \multirow{3}{*}{$10^1$}
	  & LASSO & 4 & 0 & 0 & 0 & 0 & 0 & 0 & 0   \\ \cline{2-10} %\hline
	  & SCAD & 4 & 0 & 0 & 0 & 0 & 0 & 0 & 0   \\ \cline{2-10} %\hline
	  & MCP & 4 & 0 & 0 & 0 & 0 & 0 & 0 & 0   \\ \hline %\cline{2-10}
	 \multirow{3}{*}{$10^2$}
	  & LASSO & 118 & 3 & 82 & 1 & 92 & 0 & 106 & 0   \\ \cline{2-10} %\hline
	  & SCAD & 118 & 3 & 84 & 2 & 89 & 0 & 105 & 0   \\ \cline{2-10} %\hline
	  & MCP & 119 & 4 & 81 & 3 & 90 & 1 & 105 & 1   \\ \hline %\cline{2-10}
	 \multirow{3}{*}{$10^3$}
	  & LASSO & 759 & 556 & 744 & 514 & 734 & 474 & 732 & 422   \\ \cline{2-10} %\hline
	  & SCAD & 758 & 558 & 741 & 526 & 732 & 498 & 730 & 472   \\ \cline{2-10} %\hline
	  & MCP & 758 & 560 & 743 & 544 & 734 & 518 & 731 & 513   \\ \hline %\cline{2-10}
    \end{tabular}
  \end{center}
}
\end{table*}

% q=1 %

\begin{table*}%[htb]
{ 
  \begin{center}
    \caption{ 
The result on generalization for $q=1$ and $n=100$. 
Each value indicates the average generalization error among $m=1{,}000$ trials. 
}
   \label{tab:para-est_q1_n100}
%\scriptsize
\footnotesize
%\small
    \begin{tabular}{|c|c|r|r|r|r|r|r|r|r|} \hline
	\multirow{2}{*}{$\theta_0$} & \multirow{2}{*}{Method} & \multicolumn{2}{|c|}{$r_{\mathrm{nz}}=0.2$} & \multicolumn{2}{|c|}{$r_{\mathrm{nz}}=0.4$} & \multicolumn{2}{|c|}{$r_{\mathrm{nz}}=0.6$} & \multicolumn{2}{|c|}{$r_{\mathrm{nz}}=0.8$} \\ \cline{3-10}
	 & & $d=10$ & $d=100$ & $d=10$ & $d=100$ & $d=10$ & $d=100$ & $d=10$ & $d=100$ \\ \hline
	 \multirow{3}{*}{$10^0$}
	  & LASSO & 1.03 & 1.18 & 1.05 & 1.31 & 1.06 & 1.40 & 1.07 & 1.47   \\ \cline{2-10} %\hline
	  & SCAD & 1.04 & 1.19 & 1.05 & 1.34 & 1.07 & 1.49 & 1.08 & 1.64   \\ \cline{2-10} %\hline
	  & MCP & 1.05 & 1.20 & 1.06 & 1.37 & 1.07 & 1.53 & 1.09 & 1.70   \\ \hline %\cline{2-10}
	 \multirow{3}{*}{$10^1$}
	  & LASSO & 1.07 & 1.55 & 1.09 & 2.53 & 1.11 & 3.78 & 1.11 & 4.76   \\ \cline{2-10} %\hline
	  & SCAD & 1.04 & 1.22 & 1.06 & 1.86 & 1.08 & 3.05 & 1.10 & 3.54   \\ \cline{2-10} %\hline
	  & MCP & 1.04 & 1.22 & 1.06 & 1.73 & 1.08 & 2.68 & 1.10 & 3.02   \\ \hline %\cline{2-10}
	 \multirow{3}{*}{$10^2$}
	  & LASSO & 1.07 & 21.02 & 1.09 & 99.2 & 1.10 & 225.5 & 1.11 & 321.6   \\ \cline{2-10} %\hline
	  & SCAD & 1.04 & 1.20 & 1.06 & 16.5 & 1.08 & 127.7 & 1.10 & 174.2   \\ \cline{2-10} %\hline
	  & MCP & 1.04 & 1.20 & 1.06 & 7.88 & 1.08 & 85.2 & 1.10 & 119.1   \\ \hline %\cline{2-10}
	 \multirow{3}{*}{$10^3$}
	  & LASSO & 1.07 & 1984 & 1.11 & 9762 & 1.17 & 22405 & 1.23 & 32013   \\ \cline{2-10} %\hline
	  & SCAD & 1.04 & 1.20 & 1.06 & 1557 & 1.07 & 12795 & 1.09 & 17287   \\ \cline{2-10} %\hline
	  & MCP & 1.04 & 1.20 & 1.06 & 636 & 1.08 & 8537 & 1.09 & 11803   \\ \hline %\cline{2-10}
    \end{tabular}
  \end{center}
}
\end{table*}

\begin{table*}%[htb]
{ 
  \begin{center}
    \caption{ 
The result on generalization for $q=1$ and $n=1{,}000$. 
}
   \label{tab:para-est_q1_n1000}
%\scriptsize
\footnotesize
%\small
    \begin{tabular}{|c|c|r|r|r|r|r|r|r|r|} \hline
	\multirow{2}{*}{$\theta_0$} & \multirow{2}{*}{Method} & \multicolumn{2}{|c|}{$r_{\mathrm{nz}}=0.2$} & \multicolumn{2}{|c|}{$r_{\mathrm{nz}}=0.4$} & \multicolumn{2}{|c|}{$r_{\mathrm{nz}}=0.6$} & \multicolumn{2}{|c|}{$r_{\mathrm{nz}}=0.8$} \\ \cline{3-10}
	 & & $d=10$ & $d=100$ & $d=10$ & $d=100$ & $d=10$ & $d=100$ & $d=10$ & $d=100$ \\ \hline
	 \multirow{3}{*}{$10^0$}
	  & LASSO & 1.00 & 1.02 & 1.00 & 1.04 & 1.01 & 1.05 & 1.01 & 1.06   \\ \cline{2-10} %\hline
	  & SCAD & 1.00 & 1.02 & 1.00 & 1.04 & 1.01 & 1.05 & 1.01 & 1.07   \\ \cline{2-10} %\hline
	  & MCP & 1.00 & 1.02 & 1.00 & 1.04 & 1.01 & 1.06 & 1.01 & 1.07   \\ \hline %\cline{2-10}
	 \multirow{3}{*}{$10^1$}
	  & LASSO & 1.01 & 1.05 & 1.01 & 1.08 & 1.01 & 1.09 & 1.01 & 1.10   \\ \cline{2-10} %\hline
	  & SCAD & 1.00 & 1.02 & 1.00 & 1.04 & 1.01 & 1.06 & 1.01 & 1.08   \\ \cline{2-10} %\hline
	  & MCP & 1.00 & 1.02 & 1.01 & 1.04 & 1.01 & 1.06 & 1.01 & 1.08   \\ \hline %\cline{2-10}
	 \multirow{3}{*}{$10^2$}
	  & LASSO & 1.01 & 1.05 & 1.01 & 1.08 & 1.01 & 1.09 & 1.01 & 1.10   \\ \cline{2-10} %\hline
	  & SCAD & 1.00 & 1.02 & 1.00 & 1.04 & 1.01 & 1.06 & 1.01 & 1.08   \\ \cline{2-10} %\hline
	  & MCP & 1.00 & 1.02 & 1.01 & 1.04 & 1.01 & 1.06 & 1.01 & 1.08   \\ \hline %\cline{2-10}
	 \multirow{3}{*}{$10^3$}
	  & LASSO & 1.01 & 1.06 & 1.01 & 1.13 & 1.01 & 1.22 & 1.02 & 1.33   \\ \cline{2-10} %\hline
	  & SCAD & 1.00 & 1.02 & 1.00 & 1.04 & 1.01 & 1.06 & 1.01 & 1.08   \\ \cline{2-10} %\hline
	  & MCP & 1.00 & 1.02 & 1.01 & 1.04 & 1.01 & 1.06 & 1.01 & 1.08   \\ \hline %\cline{2-10}
    \end{tabular}
  \end{center}
}
\end{table*}

% q=13/11 %

\begin{table*}%[htb]
{ 
  \begin{center}
    \caption{ 
The result on generalization for $q=13/11$ and $n=100$. 
}
   \label{tab:para-est_t10_n100}
%\scriptsize
\footnotesize
%\small
    \begin{tabular}{|c|c|r|r|r|r|r|r|r|r|} \hline
	\multirow{2}{*}{$\theta_0$} & \multirow{2}{*}{Method} & \multicolumn{2}{|c|}{$r_{\mathrm{nz}}=0.2$} & \multicolumn{2}{|c|}{$r_{\mathrm{nz}}=0.4$} & \multicolumn{2}{|c|}{$r_{\mathrm{nz}}=0.6$} & \multicolumn{2}{|c|}{$r_{\mathrm{nz}}=0.8$} \\ \cline{3-10}
	 & & $d=10$ & $d=100$ & $d=10$ & $d=100$ & $d=10$ & $d=100$ & $d=10$ & $d=100$ \\ \hline
	 \multirow{3}{*}{$10^0$}
	  & LASSO & 1.27 & 1.46 & 1.29 & 1.63 & 1.31 & 1.73 & 1.32 & 1.81   \\ \cline{2-10} %\hline
	  & SCAD & 1.27 & 1.47 & 1.29 & 1.63 & 1.31 & 1.78 & 1.33 & 1.92   \\ \cline{2-10} %\hline
	  & MCP & 1.27 & 1.47 & 1.29 & 1.65 & 1.31 & 1.81 & 1.33 & 1.97   \\ \hline %\cline{2-10}
	 \multirow{3}{*}{$10^1$}
	  & LASSO & 1.33 & 1.93 & 1.36 & 2.95 & 1.37 & 4.17 & 1.38 & 5.19   \\ \cline{2-10} %\hline
	  & SCAD & 1.29 & 1.55 & 1.31 & 2.39 & 1.34 & 3.51 & 1.36 & 4.03   \\ \cline{2-10} %\hline
	  & MCP & 1.29 & 1.55 & 1.31 & 2.24 & 1.34 & 3.12 & 1.36 & 3.51   \\ \hline %\cline{2-10}
	 \multirow{3}{*}{$10^2$}
	  & LASSO & 1.33 & 21.20 & 1.36 & 100.23 & 1.37 & 224.32 & 1.38 & 325.87   \\ \cline{2-10} %\hline
	  & SCAD & 1.28 & 1.52 & 1.31 & 17.21 & 1.33 & 127.43 & 1.36 & 177.62   \\ \cline{2-10} %\hline
	  & MCP & 1.29 & 1.52 & 1.31 & 7.86 & 1.34 & 85.82 & 1.36 & 122.05   \\ \hline %\cline{2-10}
	 \multirow{3}{*}{$10^3$}
	  & LASSO & 1.32 & 1965 & 1.37 & 9830 & 1.43 & 22282 & 1.51 & 32492   \\ \cline{2-10} %\hline
	  & SCAD & 1.28 & 1.52 & 1.31 & 1540 & 1.33 & 12921 & 1.36 & 18310   \\ \cline{2-10} %\hline
	  & MCP & 1.29 & 1.52 & 1.31 & 728 & 1.33 & 8716 & 1.36 & 12433   \\ \hline %\cline{2-10}
    \end{tabular}
  \end{center}
}
\end{table*}

\begin{table*}%[htb]
{ 
  \begin{center}
    \caption{ 
The result on generalization for $q=13/11$ and $n=1{,}000$. 
}
   \label{tab:para-est_t10_n1000}
%\scriptsize
\footnotesize
%\small
    \begin{tabular}{|c|c|r|r|r|r|r|r|r|r|} \hline
	\multirow{2}{*}{$\theta_0$} & \multirow{2}{*}{Method} & \multicolumn{2}{|c|}{$r_{\mathrm{nz}}=0.2$} & \multicolumn{2}{|c|}{$r_{\mathrm{nz}}=0.4$} & \multicolumn{2}{|c|}{$r_{\mathrm{nz}}=0.6$} & \multicolumn{2}{|c|}{$r_{\mathrm{nz}}=0.8$} \\ \cline{3-10}
	 & & $d=10$ & $d=100$ & $d=10$ & $d=100$ & $d=10$ & $d=100$ & $d=10$ & $d=100$ \\ \hline
	 \multirow{3}{*}{$10^0$}
	  & LASSO & 1.25 & 1.27 & 1.26 & 1.29 & 1.26 & 1.30 & 1.26 & 1.32   \\ \cline{2-10} %\hline
	  & SCAD & 1.25 & 1.27 & 1.26 & 1.29 & 1.26 & 1.30 & 1.26 & 1.32   \\ \cline{2-10} %\hline
	  & MCP & 1.25 & 1.27 & 1.26 & 1.29 & 1.26 & 1.31 & 1.26 & 1.33   \\ \hline %\cline{2-10}
	 \multirow{3}{*}{$10^1$}
	  & LASSO & 1.26 & 1.32 & 1.26 & 1.35 & 1.26 & 1.36 & 1.26 & 1.37   \\ \cline{2-10} %\hline
	  & SCAD & 1.25 & 1.28 & 1.26 & 1.30 & 1.26 & 1.33 & 1.26 & 1.35   \\ \cline{2-10} %\hline
	  & MCP & 1.25 & 1.28 & 1.26 & 1.30 & 1.26 & 1.33 & 1.26 & 1.35   \\ \hline %\cline{2-10}
	 \multirow{3}{*}{$10^2$}
	  & LASSO & 1.26 & 1.32 & 1.26 & 1.35 & 1.26 & 1.36 & 1.26 & 1.37   \\ \cline{2-10} %\hline
	  & SCAD & 1.25 & 1.28 & 1.26 & 1.30 & 1.26 & 1.33 & 1.26 & 1.35   \\ \cline{2-10} %\hline
	  & MCP & 1.25 & 1.28 & 1.26 & 1.30 & 1.26 & 1.33 & 1.26 & 1.35   \\ \hline %\cline{2-10}
	 \multirow{3}{*}{$10^3$}
	  & LASSO & 1.26 & 1.32 & 1.26 & 1.39 & 1.27 & 1.49 & 1.27 & 1.60   \\ \cline{2-10} %\hline
	  & SCAD & 1.25 & 1.28 & 1.26 & 1.30 & 1.26 & 1.33 & 1.26 & 1.35   \\ \cline{2-10} %\hline
	  & MCP & 1.25 & 1.28 & 1.26 & 1.30 & 1.26 & 1.33 & 1.26 & 1.35   \\ \hline %\cline{2-10}
    \end{tabular}
  \end{center}
}
\end{table*}

% q=3/2 %

\begin{table*}%[htb]
{ 
  \begin{center}
    \caption{ 
The result on generalization for $q=3/2$ and $n=100$. 
}
   \label{tab:para-est_t3_n100}
%\scriptsize
\footnotesize
%\small
    \begin{tabular}{|c|c|r|r|r|r|r|r|r|r|} \hline
	\multirow{2}{*}{$\theta_0$} & \multirow{2}{*}{Method} & \multicolumn{2}{|c|}{$r_{\mathrm{nz}}=0.2$} & \multicolumn{2}{|c|}{$r_{\mathrm{nz}}=0.4$} & \multicolumn{2}{|c|}{$r_{\mathrm{nz}}=0.6$} & \multicolumn{2}{|c|}{$r_{\mathrm{nz}}=0.8$} \\ \cline{3-10}
	 & & $d=10$ & $d=100$ & $d=10$ & $d=100$ & $d=10$ & $d=100$ & $d=10$ & $d=100$ \\ \hline
	 \multirow{3}{*}{$10^0$}
	  & LASSO & 2.87 & 3.29 & 2.89 & 3.50 & 2.91 & 3.68 & 2.93 & 3.83   \\ \cline{2-10} %\hline
	  & SCAD & 2.87 & 3.29 & 2.89 & 3.50 & 2.91 & 3.67 & 2.93 & 3.83   \\ \cline{2-10} %\hline
	  & MCP & 2.88 & 3.32 & 2.89 & 3.51 & 2.91 & 3.69 & 2.93 & 3.87   \\ \hline %\cline{2-10}
	 \multirow{3}{*}{$10^1$}
	  & LASSO & 3.00 & 4.65 & 3.07 & 5.82 & 3.11 & 7.05 & 3.13 & 8.09   \\ \cline{2-10} %\hline
	  & SCAD & 2.94 & 4.47 & 3.01 & 6.04 & 3.07 & 6.90 & 3.11 & 7.44   \\ \cline{2-10} %\hline
	  & MCP & 2.93 & 5.50 & 3.01 & 5.93 & 3.06 & 6.58 & 3.10 & 6.95   \\ \hline %\cline{2-10}
	 \multirow{3}{*}{$10^2$}
	  & LASSO & 3.01 & 23.33 & 3.08 & 103.52 & 3.11 & 229.03 & 3.14 & 331.08   \\ \cline{2-10} %\hline
	  & SCAD & 2.91 & 3.72 & 2.96 & 46.11 & 3.02 & 133.22 & 3.08 & 180.93   \\ \cline{2-10} %\hline
	  & MCP & 2.91 & 3.72 & 2.97 & 11.23 & 3.02 & 91.05 & 3.09 & 125.76   \\ \hline %\cline{2-10}
	 \multirow{3}{*}{$10^3$}
	  & LASSO & 3.00 & 1973 & 3.08 & 9900 & 3.16 & 22441 & 3.26 & 32653   \\ \cline{2-10} %\hline
	  & SCAD & 2.91 & 3.72 & 2.96 & 1523 & 3.02 & 12689 & 3.08 & 18575   \\ \cline{2-10} %\hline
	  & MCP & 2.91 & 3.72 & 2.97 & 677 & 3.02 & 8584 & 3.08 & 12104   \\ \hline %\cline{2-10}
    \end{tabular}
  \end{center}
}
\end{table*}

\begin{table*}%[htb]
{ 
  \begin{center}
    \caption{ 
The result on generalization for $q=3/2$ and $n=1{,}000$. 
}
   \label{tab:para-est_t3_n1000}
%\scriptsize
\footnotesize
%\small
    \begin{tabular}{|c|c|r|r|r|r|r|r|r|r|} \hline
	\multirow{2}{*}{$\theta_0$} & \multirow{2}{*}{Method} & \multicolumn{2}{|c|}{$r_{\mathrm{nz}}=0.2$} & \multicolumn{2}{|c|}{$r_{\mathrm{nz}}=0.4$} & \multicolumn{2}{|c|}{$r_{\mathrm{nz}}=0.6$} & \multicolumn{2}{|c|}{$r_{\mathrm{nz}}=0.8$} \\ \cline{3-10}
	 & & $d=10$ & $d=100$ & $d=10$ & $d=100$ & $d=10$ & $d=100$ & $d=10$ & $d=100$ \\ \hline
	 \multirow{3}{*}{$10^0$}
	  & LASSO & 2.92 & 2.92 & 2.92 & 2.94 & 2.92 & 2.96 & 2.93 & 2.98   \\ \cline{2-10} %\hline
	  & SCAD & 2.92 & 2.92 & 2.92 & 2.94 & 2.92 & 2.96 & 2.93 & 2.98   \\ \cline{2-10} %\hline
	  & MCP & 2.92 & 2.92 & 2.92 & 2.94 & 2.92 & 2.96 & 2.93 & 2.98   \\ \hline %\cline{2-10}
	 \multirow{3}{*}{$10^1$}
	  & LASSO & 2.93 & 3.05 & 2.94 & 3.12 & 2.95 & 3.17 & 2.95 & 3.19   \\ \cline{2-10} %\hline
	  & SCAD & 2.93 & 2.99 & 2.94 & 3.05 & 2.94 & 3.11 & 2.95 & 3.16   \\ \cline{2-10} %\hline
	  & MCP & 2.93 & 2.99 & 2.94 & 3.05 & 2.94 & 3.11 & 2.95 & 3.16   \\ \hline %\cline{2-10}
	 \multirow{3}{*}{$10^2$}
	  & LASSO & 2.93 & 3.05 & 2.94 & 3.12 & 2.95 & 3.17 & 2.95 & 3.19   \\ \cline{2-10} %\hline
	  & SCAD & 2.92 & 2.96 & 2.93 & 3.02 & 2.94 & 3.08 & 2.94 & 3.14   \\ \cline{2-10} %\hline
	  & MCP & 2.92 & 2.96 & 2.93 & 3.02 & 2.94 & 3.08 & 2.94 & 3.14   \\ \hline %\cline{2-10}
	 \multirow{3}{*}{$10^3$}
	  & LASSO & 2.93 & 3.05 & 2.94 & 3.14 & 2.95 & 3.26 & 2.96 & 3.40   \\ \cline{2-10} %\hline
	  & SCAD & 2.92 & 2.96 & 2.93 & 3.02 & 2.94 & 3.08 & 2.94 & 3.14   \\ \cline{2-10} %\hline
	  & MCP & 2.93 & 2.96 & 2.93 & 3.02 & 2.94 & 3.08 & 2.94 & 3.14   \\ \hline %\cline{2-10}
    \end{tabular}
  \end{center}
}
\end{table*}

% q=5/3 %

\begin{table*}%[htb]
{ 
  \begin{center}
    \caption{ 
The result on generalization for $q=5/3$ and $n=100$. 
}
   \label{tab:para-est_t2_n100}
%\scriptsize
\footnotesize
%\small
    \begin{tabular}{|c|c|r|r|r|r|r|r|r|r|} \hline
	\multirow{2}{*}{$\theta_0$} & \multirow{2}{*}{Method} & \multicolumn{2}{|c|}{$r_{\mathrm{nz}}=0.2$} & \multicolumn{2}{|c|}{$r_{\mathrm{nz}}=0.4$} & \multicolumn{2}{|c|}{$r_{\mathrm{nz}}=0.6$} & \multicolumn{2}{|c|}{$r_{\mathrm{nz}}=0.8$} \\ \cline{3-10}
	 & & $d=10$ & $d=100$ & $d=10$ & $d=100$ & $d=10$ & $d=100$ & $d=10$ & $d=100$ \\ \hline
	 \multirow{3}{*}{$10^0$}
	  & LASSO & 10.68 & 14.07 & 10.71 & 14.26 & 10.72 & 14.46 & 10.74 & 14.65   \\ \cline{2-10} %\hline
	  & SCAD & 10.68 & 14.07 & 10.71 & 14.26 & 10.72 & 14.47 & 10.74 & 14.66   \\ \cline{2-10} %\hline
	  & MCP & 10.69 & 14.09 & 10.71 & 14.29 & 10.72 & 14.49 & 10.74 & 14.69   \\ \hline %\cline{2-10}
	 \multirow{3}{*}{$10^1$}
	  & LASSO & 11.04 & 17.72 & 11.23 & 19.45 & 11.32 & 21.00 & 11.40 & 22.28   \\ \cline{2-10} %\hline
	  & SCAD & 10.97 & 18.85 & 11.18 & 21.89 & 11.31 & 22.62 & 11.41 & 25.29   \\ \cline{2-10} %\hline
	  & MCP & 10.97 & 19.12 & 11.19 & 22.11 & 11.33 & 23.14 & 11.42 & 24.98   \\ \hline %\cline{2-10}
	 \multirow{3}{*}{$10^2$}
	  & LASSO & 11.32 & 37.51 & 11.64 & 120.93 & 11.74 & 245.48 & 11.78 & 346.97   \\ \cline{2-10} %\hline
	  & SCAD & 11.00 & 18.13 & 11.31 & 38.66 & 11.51 & 149.72 & 11.65 & 199.80   \\ \cline{2-10} %\hline
	  & MCP & 11.00 & 18.05 & 11.30 & 27.55 & 11.51 & 107.95 & 11.66 & 144.35   \\ \hline %\cline{2-10}
	 \multirow{3}{*}{$10^3$}
	  & LASSO & 11.31 & 1978.49 & 11.65 & 10079 & 11.77 & 22466 & 11.89 & 32608   \\ \cline{2-10} %\hline
	  & SCAD & 10.89 & 16.41 & 11.18 & 1627 & 11.39 & 12520 & 11.60 & 17490   \\ \cline{2-10} %\hline
	  & MCP & 10.90 & 16.41 & 11.18 & 664 & 11.39 & 8513 & 11.61 & 12066   \\ \hline %\cline{2-10}
    \end{tabular}
  \end{center}
}
\end{table*}

\begin{table*}%[htb]
{ 
  \begin{center}
    \caption{ 
The result on generalization for $q=5/3$ and $n=1{,}000$. 
}
   \label{tab:para-est_t2_n1000}
%\scriptsize
\footnotesize
%\small
    \begin{tabular}{|c|c|r|r|r|r|r|r|r|r|} \hline
	\multirow{2}{*}{$\theta_0$} & \multirow{2}{*}{Method} & \multicolumn{2}{|c|}{$r_{\mathrm{nz}}=0.2$} & \multicolumn{2}{|c|}{$r_{\mathrm{nz}}=0.4$} & \multicolumn{2}{|c|}{$r_{\mathrm{nz}}=0.6$} & \multicolumn{2}{|c|}{$r_{\mathrm{nz}}=0.8$} \\ \cline{3-10}
	 & & $d=10$ & $d=100$ & $d=10$ & $d=100$ & $d=10$ & $d=100$ & $d=10$ & $d=100$ \\ \hline
	 \multirow{3}{*}{$10^0$}
	  & LASSO & 25.40 & 12.63 & 25.41 & 12.65 & 25.41 & 12.67 & 25.41 & 12.69   \\ \cline{2-10} %\hline
	  & SCAD & 25.40 & 12.63 & 25.41 & 12.65 & 25.41 & 12.67 & 25.41 & 12.69   \\ \cline{2-10} %\hline
	  & MCP & 25.40 & 12.63 & 25.41 & 12.65 & 25.41 & 12.67 & 25.41 & 12.69   \\ \hline %\cline{2-10}
	 \multirow{3}{*}{$10^1$}
	  & LASSO & 25.45 & 13.08 & 25.48 & 13.32 & 25.49 & 13.49 & 25.50 & 13.59   \\ \cline{2-10} %\hline
	  & SCAD & 25.45 & 13.07 & 25.48 & 13.34 & 25.50 & 13.52 & 25.50 & 13.63   \\ \cline{2-10} %\hline
	  & MCP & 25.45 & 13.08 & 25.47 & 13.35 & 25.50 & 13.53 & 25.51 & 13.64   \\ \hline %\cline{2-10}
	 \multirow{3}{*}{$10^2$}
	  & LASSO & 25.48 & 13.25 & 25.53 & 13.57 & 25.57 & 13.75 & 25.58 & 13.84   \\ \cline{2-10} %\hline
	  & SCAD & 25.45 & 12.90 & 25.49 & 13.17 & 25.54 & 13.42 & 25.57 & 13.66   \\ \cline{2-10} %\hline
	  & MCP & 25.45 & 12.90 & 25.50 & 13.18 & 25.54 & 13.41 & 25.57 & 13.66   \\ \hline %\cline{2-10}
	 \multirow{3}{*}{$10^3$}
	  & LASSO & 25.52 & 13.25 & 25.56 & 13.57 & 25.58 & 13.79 & 25.65 & 14.00   \\ \cline{2-10} %\hline
	  & SCAD & 25.44 & 12.85 & 25.47 & 13.11 & 25.50 & 13.37 & 25.59 & 13.63   \\ \cline{2-10} %\hline
	  & MCP & 25.43 & 12.85 & 25.47 & 13.11 & 25.50 & 13.37 & 25.59 & 13.63   \\ \hline %\cline{2-10}
    \end{tabular}
  \end{center}
}
\end{table*}

\section{Conclusion}
\label{sec:conclusion}

We proposed regularization methods for $q$-normal linear models based on the $L_q$-likelihood. 
The proposed methods coincide with the ordinary regularization methods. 
Our methods perform well for slightly heavy-tailed errors ($1 \leq q < 2$) in terms of model selection and generalization. 
Moreover, they work well in terms of model selection for heavy-tailed errors ($2 \leq q < 2.5$). 
A theoretical analysis of the proposed methods is left to future work.

\bibliographystyle{plain}
\bibliography{arxiv2020qnormal_v1}

\end{document}